# Modelling Optomechanical Responses in Optical Tweezers Beyond Paraxial Limits

Moosung Lee[1,2,†,*], Tobias Hanke[1,2,†], Sara Launer[1,2], and Sungkun Hong[1,2*]

[1] *Institute for Functional Matter and Quantum Technologies, University of Stuttgart, 70569 Stuttgart, Germany;*

[2] *Center for Integrated Quantum Science and Technology, University of Stuttgart, 70569 Stuttgart, Germany;*

[†] These authors equally contributed to the work.

* *Corresponding authors: M.L ($^{moosung.lee@fmq.uni-stuttgart.de}$), S. H ($^{sungkun.hong@fmq.uni-stuttgart.de}$)*

**E-mail addresses**

Moosung Lee: moosung.lee@fmq.uni-stuttgart.de

Tobias Hanke: tobiashanke1998@gmail.com

Sara Launer: launersara@gmail.com

Sungkun Hong: sungkun.hong@fmq.uni-stuttgart.de

**ABSTRACT**

Optically levitated dielectric nanoparticles have become valuable tools for precision sensing and quantum optomechanical experiments. To predict the dynamic properties of a particle trapped in an optical tweezer with high fidelity, a tool is needed to compute the particle's response to the given optical field accurately. Here, we utilise a numerical solution of the three-dimensional trapping light to accurately simulate optical tweezers and predict key optomechanical parameters. By controlling the numerical aperture and measuring the the particle's oscillation frequencies in the trap, we validate the accuracy of our method. We foresee broad applications of this method in the field of levitodynamics, where precise characterisation of optical tweezers is essential for estimating parameters ranging from motional frequencies to scattering responses of the particle with various dielectric properties.

# INTRODUCTION

Since their invention[1], optical tweezers have become indispensable tools in various scientific fields, ranging from biophysics[2] to ultracold atom experiments[3]. More recently, levitating dielectric nanoparticles, particularly in vacuum[4], have found exciting applications in levitated optomechanics[5]. Isolating optically trapped nanoparticles in a high vacuum provides exceptional quality factors, facilitating quantum-limited control[6–8] necessary for macroscopic quantum tests[9] and precision sensing[10–12].

To evaluate systematic performance in levitated optomechanics, rigorous modelling of optical trapping is necessary. This process involves accurately predicting the light-matter interactions between the trapping light and the trapped nanoparticle, as well as the particle's subsequent motion within optical tweezers. The key parameter determining the particle's dynamics in optical tweezers is its oscillation frequency, also called trap frequency. This frequency is influenced by multiple factors such as the size and refractive index of the trapped particle[13] and laser power and polarisation[14]. Among these parameters, numerical aperture (NA) plays a particularly unique role in determining the size of the focused beam[5] and, thus, the shape of the trap potential in optical tweezers. The paraxial approximation has been commonly used to model the effect of the NA on optical traps[13,15]. However, the approximation inherent in the model limits its accuracy, hindering precise characterisation and analysis of optical tweezers in experiments.

Here we introduce full-field modelling of the trapping light to accurately analyse the characteristics of an optical tweezer. Based on the vectoral angular spectrum method, our approach allows us to numerically deduce the solution of the optical tweezer field in 3D. We validate our model by experimentally controlling the focusing NA of an optical tweezer and measuring the particle's motion in the trap. We found that our method based on the full-field modelling, in stark contrast to the paraxial approximation, could precisely predict the experimentally measured oscillation frequencies of the particle depending on the NA. We propose that the demonstrated method could also be used to estimate other important optomechanical parameters, including the scattering response of the particle and the magnitude of photon recoil heating.

## Results

**Experimental Setup**

The schematic of our experimental setup is illustrated in Fig. 1. The setup includes an iris diaphragm positioned at the pupil plane to control the focusing NA of the trapping light [Fig. 1(a)]. Specifically, a continuous-wave infrared laser (wavelength λ = 1064 nm; Azurlight Systems) was spatially filtered using a polarisation-maintaining single-mode fibre and then expanded to create a plane wave [Fig. 1(b)]. An iris diaphragm positioned at the pupil plane cropped the incoming light, with the radius of the iris aperture determining the system NA. The iris aperture was demagnified by a factor of 2.5 using two 4-*f* telescopic arrays so that, when fully open, the largest beam size at the pupil plane matched the entrance pupil of the objective lens used (MPLN 100X; NA = 0.9; Olympus Inc.). To estimate the NA and pupil intensity of the system, the image of the incident beam was recorded by a monitor camera (1800 U-507M mono, Allied Vision Inc.). To determine the pupil NA, the intensity image was segmented using Otsu's method[16] and fitted to a Gaussian distribution.

We trapped 142-nm-diameter silica nanoparticles (microParticles GmbH) as a reference sample for optical trapping, owing to their exceptional stability in a vacuum and known material properties. The particle was optically trapped at 0.4 millibar and monitored through an additional 4-*f* telescope positioned at the side view. The light backscattered by a trapped nanoparticle was redirected by a polarising beam splitter and a quarter-wave plate and was detected with a balanced photodetector to measure the motion of the trapped particle and, thus, its oscillation frequencies along three spatial axes.

**Analysis of point spread function**

To analyse and compare the accuracy of the beam models, we first measured the 3D point spread function (PSF) as a function of the system's NA. To achieve this, we placed a mirror equipped with a linear piezoelectric stage (PDX1/M; Thorlabs Inc.) at the focal plane and scanned it to acquire the 3D PSF [Fig. 2(a)]. We then compared the obtained 3D PSF with the PSFs predicted from two different theoretical models: the conventional scalar paraxial approximation[17] and our method based on the vectoral angular spectrum method[18].

First, we employed a Gaussian beam, a conventional and simplistic method used to describe the optical tweezers[17]. In the model, the light propagation is based on the scalar paraxial approximation, and the intensity is given by:

$$I(\mathbf{r}_\perp, z) = \frac{2P}{\pi w_0^2} \frac{1}{1+\{\lambda z/(\pi w_0^2)\}^2} \exp\left(-\frac{2\mathbf{r}_\perp^2}{w_0^2[1+\{\lambda z/(\pi w_0^2)\}^2]}\right), \quad (1)$$

where $P$ is the laser power and $w_0$ is the beam waist parameter. According to the theory of Fourier optics[17], the relationship between $w_0$ and NA is inversely proportional, such that $w_0 = \beta \cdot \lambda / (\pi \cdot \text{NA})$. Since the Gaussian beam model does not impose any constraint on $w_0$, a system-specific correction coefficient, $\beta$, is determined empirically. When $\beta$ is assumed to be 1 for convenience[15,17], the spot sizes of theoretical PSFs were found to be significantly smaller than the experimental ones [Fig. 2(a)].

To address this discrepancy, we fitted the experimentally measured FWHMs to those predicted by the Gaussian beam, with $\beta$ being a fitting parameter. We found that the lateral FWHMs of the Gaussian beam, ($\Delta x = \sqrt{2\ln 2}\, w_0 \approx 0.375\beta \cdot \lambda/\text{NA}$) were in good agreement when $\beta = 1.4107 \approx \sqrt{2}$ [Fig. 2(b)]. However, when we used the same value of $\beta$ to estimate the axial FWHMs, $\Delta z = 2\pi \cdot w_0^2/\lambda \approx 0.637\beta^2 \cdot \lambda / \text{NA}^2$, a deviation from the experimental results was observed [Fig. 2(c)]. This discrepancy is primarily attributed to the inherent limitations of the paraxial approximation, limiting the accuracy in the axial intensity profile.

We then employed the vectoral angular spectrum method[18] to model the optical tweezers more accurately. In cylindrical coordinates, the complex amplitude of focused monochromatic light is given by the following relationship:

$$\mathbf{E}(\rho, \varphi, z) \propto \int_0^{\theta_{max}} \int_0^{2\pi} \mathbf{E}_\infty(\theta, \phi) e^{ikz\cos\theta} e^{ik\rho\sin\theta\cos(\phi-\varphi)} \sin\theta \, d\phi d\theta \quad (2)$$

where $\mathbf{E}_\infty(\theta, \phi)$ is the far-field amplitude of the incident light, $k = 2\pi/\lambda$ is the spatial wavenumber of the light, and $\theta_{max} = \sin^{-1} \text{NA}$ is the maximal acceptance angle determined by the system NA. Assuming circular symmetry in the amplitude distribution of the incoming light, the solution to the *x*-polarised focal light field is given by[18]:

$$\mathbf{E}_x(\rho, \phi, z) \propto \begin{bmatrix} I_{00}(\rho, z) + I_{02} \cos 2\phi \\ I_{02}(\rho, z) \sin 2\phi \\ -2i I_{01}(\rho, z) \cos \phi \end{bmatrix}, \quad (3)$$

where

$$\begin{bmatrix} I_{00}(\rho, z) \\ I_{01}(\rho, z) \\ I_{02}(\rho, z) \end{bmatrix} = \int_0^{\theta_{NA}} \sqrt{f_w(\theta) \cdot \cos \theta} \sin \theta \begin{bmatrix} (1 + \cos \theta) J_0(k\rho \sin \theta) \\ (\sin \theta) J_1(k\rho \sin \theta) \\ (1 - \cos \theta) J_2(k\rho \sin \theta) \end{bmatrix} e^{ikz \cos \theta} d\theta. \quad (4)$$

Here, $J_n$ is the *n*-th order Bessel function. $f_w(\theta) \propto \exp\left(-\frac{\sin^2 \theta}{w_{NA}^2}\right)$ represents the experimentally measured pupil intensity. In our experiment, $w_{NA}$, the Gaussian blurriness at the pupil plane, was estimated through image analysis and determined to be $0.784 \pm 0.007$ (mean ± standard error). A similar relation can be derived for *y*-polarised light to simulate arbitrary polarised light.

Using this model, we simulated the NA-dependent 3D PSFs, considering the circular polarisation and $f_w(\theta)$ measured from the monitor camera. As depicted in Fig. 2(a), the simulated PSFs best agreed with the experimental results. To quantitatively observe this agreement, we calculated the lateral and axial FWHMs as a function of NA [Figs. 2(b, c)]. A comparison between the experimental data and the numerical estimates of the FWHMs demonstrates that our rigorous simulation method provided the closest alignment. In particular, our model yielded axial FWHMs which were the most consistent with the experimental measurements. To further quantify this agreement, we estimated the ratio between the experimentally measured and the theoretically predicted axial FWHM [Fig. 2(d)]. This ratio is ideally 1, and our results show that this value was closest to 1 using our method, despite experimental errors caused by imperfections in the piezoelectric stages, as well as unwanted multiple reflections and vignetting of the focused light. Taken together, our experimental PSF analysis validates the accuracy of our beam modelling based on the vectoral angular spectrum method.

**Analysis of trapping frequency**

Building on the PSF analysis, we next investigated whether our method based on the angular spectrum method also improves the accuracy of estimating the trap frequencies of optically trapped nanoparticles along the axial ($\Omega_z/2\pi$) and lateral ($\Omega_x/2\pi$) directions. Since the particle diameter is approximately seven times smaller than the wavelength of light, the system operates in the Rayleigh regime[13]. In this regime, the trap stiffness along the $j$-th axis is governed by the gradient force of the optical tweezers, given by:

$$\kappa_j = -\frac{\alpha}{2\varepsilon_0 c}\frac{\partial^2 I}{\partial x_j^2} = -\frac{3V}{2c}\left(\frac{n^2-1}{n^2+2}\right)\frac{\partial^2 I}{\partial x_j^2}, \tag{5}$$

where $\alpha = 3V\varepsilon_0(n^2-1)/(n^2+2)$ is the polarizability with $V$ being the particle's volume, $\varepsilon_0$ the vacuum permittivity, $c$ the speed of light, $n$ the particle's refractive index and $I$ the laser intensity at the particle location. When the trapped particle is located near the centre of the beam, the angular trapping frequencies along the $j$-th axis, $\Omega_j = \sqrt{\kappa_j/m}$ ($m$: the mass of the particle), can be determined as follows:

$$\Omega_j = \sqrt{\frac{12P}{\pi c\rho}\left(\frac{n^2-1}{n^2+2}\right)}\frac{1}{w_x w_j}, \tag{6}$$

where $P$ is the laser power, $\rho$ is the particle's density, and $w_j$ is the beam width parameter of the 3D PSF along the $j$-th axis, fitted with a Gaussian distribution, $I(\mathbf{r}) \approx \frac{2P}{\pi w_x^2}\exp\left[-2\left(\frac{x^2+y^2}{w_x^2}+\frac{z^2}{w_z^2}\right)\right]$. With Eq. (1), the paraxial approximation further simplifies Eq. (6) into polynomials of NA:

$$\Omega_x = \sqrt{\frac{12\pi^3 P}{c\rho}\left(\frac{n^2-1}{n^2+2}\right)}\frac{NA^2}{\beta^2\lambda^2}, \quad \Omega_z = \sqrt{\frac{6\pi^3 P}{c\rho}\left(\frac{n^2-1}{n^2+2}\right)}\frac{NA^3}{\beta^3\lambda^2}. \tag{7}$$

Equations (6) and (7) suggest that the ratio $\Omega_z/\Omega_x = w_x/w_z = \Delta x/\Delta z$ can serve as a calibration parameter independent of the selected material or laser power, directly showing the ratios of the lateral and axial FWHMs and, thus, the geometric shape of the optical potential. For the paraxial approximation, this ratio indicates a linear relation with NA as follows:

$$\frac{\Omega_z}{\Omega_x} = \frac{\text{NA}}{\sqrt{2} \cdot \beta}. \tag{8}$$

To validate the theoretical predictions by the two models, we trapped a silica nanoparticle in our optical tweezer setup and measured the particle's oscillation frequencies $\Omega_z$ and $\Omega_x$ at different NA's ranging from 0.5 to 0.75 [Fig. 3]. This range was chosen to prevent particle loss at lower NA and to avoid unwanted aberrations and vignetting at higher NA. We repeated the experiments with three different silica nanoparticles to further confirm the reproducibility of the results. Figure 3(a) shows the displacement power spectral density measured with one of the particles with three different NA values, where the particle's motional frequencies along the lateral (z) and axial (x) axes are clearly visible. Also noticeable are the higher harmonics and mixed frequency components, which are mainly caused by the anharmonicity of the optical trap[19]. As expected, both $\Omega_x$ and $\Omega_z$ increased with increasing NA.

We first analysed the relationship between the NA and the frequency ratio $\Omega_z/\Omega_x$. As mentioned above, this parameter can be calculated from the models solely based on the NA and the profile distribution of the input field. This makes it a particularly convenient parameter for experimentally verifying the accuracy of the model without calibration of the laser power or the material properties of the particle. The results showed a significant discrepancy when using the Gaussian beam model [Fig. 3(b)]. In contrast, our full-field simulation accurately predicted the ratios across the tested range of NA. The results strongly suggest our model's superiority in accurately analysing the tweezer field and predicting key trapping parameters.

To further validate the accuracy of the model, it is necessary to assess how precisely the absolute values of the particle frequencies can be predicted. This can be achieved by specifying the laser power transmitted to the tweezers and the density and refractive index of the particle. In our experiment, the particle's material properties (density $\rho$ = 1850 kg m$^{-3}$; refractive index $n$ = 1.4496) were available from the nominal specifications provided by the vendor. However, the exact input laser power could not be determined due to the inaccuracy of the power meter used in the experiment.

In light of these limitations, we proceeded to calculate the expected tweezer beam powers at the focal plane from the measured frequencies of the particle based on the models (Eq. (6)):

$$P_{simul} = \frac{\pi c \rho}{12}\left(\frac{n^2+2}{n^2-1}\right)\left(w_x w_j \Omega_j\right)^2, \quad (9)$$

where $w_x$ and $w_z$ were calculated from our model for a given NA. We then assessed whether these values followed the NA-dependent tweezer beam power according to our experimental setting. In our setup, NA was controlled by clipping the fixed input laser beam with an iris, resulting in the following equation:

$$P_{exp} = 2\pi \int_0^{\theta_{NA}} \sin\theta \cdot f_w(\theta) d\theta = P_0 \cdot [1 - \exp(-\text{NA}^2/w_{NA}^2)], \quad (10)$$

where $P_0$ is the only undetermined free parameter.

When fitting the powers extracted from the measured lateral oscillation frequencies of the particle ($\Omega_x$) to the above equation (Eq. (10)), we obtained an excellent fit result with $R^2 = 0.912$ and $P_0 = 1291 \pm 20$ mW [Fig. 3(c)]. We also performed the fitting based on the powers extracted from measured axial frequencies ($\Omega_z$) and obtained similarly good results with $R^2 = 0.891$ and $P_0 = 1300 \pm 22$ mW. These results reaffirm the validity and reliability of our model based on the vectoral angular spectrum method.

**Prediction of trap frequency and recoil heating**

The particle's scattering power and recoil heating for a given trapping laser power are crucial parameters determining the fundamental limits in the particle's position measurement sensitivity and subsequent feedback control efficiency[20,21]. Building on the validity of our model demonstrated above, we calculate the expected recoil heating rates for different NA and show how they deviate from the prediction given by the Gaussian beam model. In the Rayleigh regime, the recoil heating rate along the $j$-th axis, $\Gamma_{\text{recoil}, j}$, is expressed by[20,22]:

$$\Gamma_{\text{recoil},j} = \frac{1}{5}\frac{P_{scat}}{mc^2}\frac{\omega_0}{\Omega_j} = \sqrt{\left(\frac{1}{\pi c}\cdot\frac{n^2-1}{n^2+2}\right)^3\frac{3P}{\rho}}\cdot\frac{k^4\omega_0 V}{10}\frac{w_j}{w_x}, \quad P_{scat} = \frac{3V^2 k^4 P}{\pi^2 w_x^2}\left(\frac{n^2-1}{n^2+2}\right)^2, \tag{11}$$

where $m$ is the particle mass, $\omega_0$ the angular frequency of the optical tweezers, and $P_{scat}$ is the scattering power depending on the incident focal power $P$. Assuming $P = 250$ mW, we predicted the trap frequencies, scattering power, and recoil heating rate [Fig. 4]. The results indicate that, compared to the full-field simulation, the paraxial approximation overestimates both the absolute trap frequencies [Fig. 4(a)] and the scattering power [Fig. 4(b)]. The higher the NA is, the trend is starker. For the recoil heating rates, both the paraxial approximation and our method predict the same values, since it is independent of the beam waist parameters [Fig. 4(c)]. By contrast, the axial recoil heating rate, which is proportional to $w_z/w_x$, exhibits a significant difference between the beam models.

The parameters predicted in our study may provide reference data for various levitodynamic experiments. For example, we note that the trap frequencies predicted by the full-field simulation agree well with previous studies[7,14]. The scattering power and the photon recoil rate, which were previously predicted using the paraxial approximation[7,20], can also be compared with the full-field simulation. The codes used in our studies have been made publicly available online to facilitate further use and validation under diverse experimental conditions (see Code and Data Availability).

## Discussion

In summary, we have presented a rigorous approach to modelling 3D optical tweezers, with the aim of systematically calibrating and predicting trap parameters relevant to levitated optomechanics. Our method involves utilising an iris diaphragm to control NA and applying the vectoral angular spectrum method to model the focused light intensity. This approach enabled accurate predictions of the lateral and axial oscillation frequencies of optically trapped nanoparticles and precise recalibration of the incident laser power. The calibrated measurements were additionally used to predict the scattering power and the photon recoil rate. Understanding these parameters will be indispensable for predicting the feasibility of quantum-limited detection and control of a particle's motion[21,23]. This model is generalisable and available online, and we anticipate its wide applicability in diverse research directions in the field of levitodynamics.

In our experiment, noticeable discrepancies between experimental and theoretical results were observed in the PSF analysis (Fig. 2). Considering the limited mechanical stability and optical imperfections of our PSF measurement, this discrepancy was expected. In contrast, our model's predictions on the particle's trap frequencies showed remarkable agreement with the measurements (Fig. 3). This suggests that the analysis of the measured particle frequencies based on our rigorous beam modelling can provide more accurate information about the 3D spatial resolution of a microscopy system, free from the stability issues of the imaging system.

The study presented here has only considered the response of Rayleigh particles within a limited NA range. We expect that several refinements to our framework can extend the scope of our model. For instance, to calculate trap frequencies of particles beyond the Rayleigh regime, methods for solving multiple light scattering problems, such as Mie theory, can be employed[24,25]. In the experiment, the range of focusing NA was kept below 0.8 to avoid pronounced aberration effects near the specified NA of the objective (0.9), which our current theoretical model does not consider. This effect can be considered in the future work. We also note that a spatial light modulator can also be used to experimentally compensate for the aberration[26–28].

Our work suggests that, once the focal laser power of the system is calibrated, our setup can determine the polarizability-to-density ratio $(1/\rho) \cdot (n^2 - 1) / (n^2 + 2)$ of an unknown nanoparticle. If the refractive index or the density of the particle can be measured independently, our setup, combined with the full-field modelling, can be further extended to unambiguously determine the particle's material origin. This can be achieved in several ways. For example, the mass of levitated particles could be determined by light-induced active control[29]. The refractive index of a levitated particle could also be retrieved using quantitative phase imaging techniques[30,31]. These techniques would enable the quantitative identification and characterisation of levitated unconventional nanomaterials, including nano-diamonds with single spin defects[32,33], rare-earth doped nanocrystals[34,35], birefringent nanomaterials[36], and other high-refractive-index materials[37]—each a promising candidate for next-generation sensing[38] and quantum[39] applications in levitodynamics.

**Acknowledgements**

We thank Minshik Kwon for his support in the early stages of this research. This research was supported by the Center for Integrated Quantum Science and Technology (IQST, Johannes-Kepler Grant) through the Carl Zeiss Foundation and by the Ministry of Science, Research and Arts of Baden-Württemberg, Germany. M.L. acknowledges the support funded by the Alexander Von Humboldt Foundation.

**Conflict of interest**

The authors declare no conflict of interest.

**Author contributions**

M.L., and T.H. performed and analysed the experiments and provided analytical tools. S.L. supported the experiments. M.L., and S.H. supervised the project. All authors wrote the manuscript.

**Data and code Availability**

The representative dataset and codes are available at https://github.com/moosunglee/NAeffect.

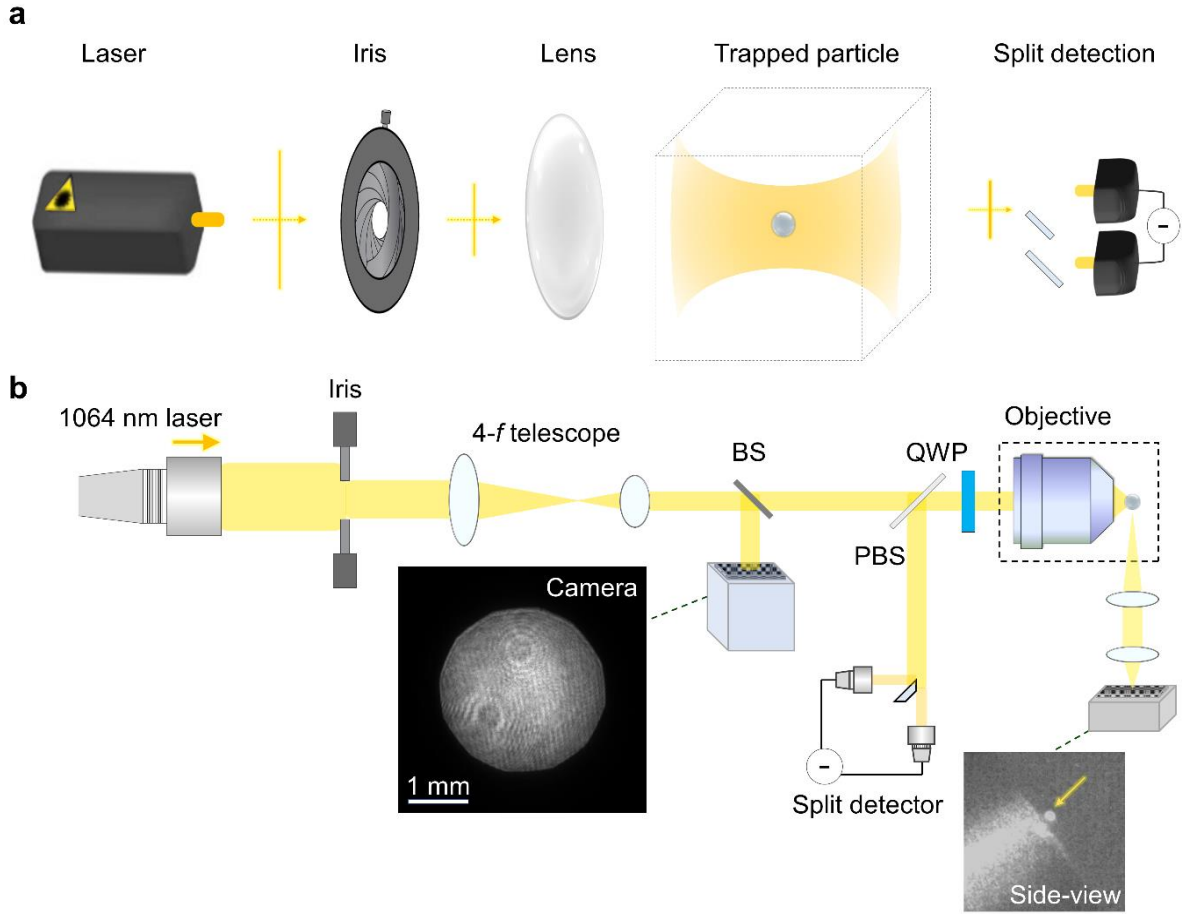

**Figure 1 | Experimental setup**

**a,** Setup schematic. The incident plane wave of a monochromatic laser is cropped by an iris diaphragm to adjust the numerical aperture (NA) at the pupil plane. The motion of the trapped particle is then recorded using split detection of the scattered light. **b,** Experimental setup in detail. A plane wave, cropped by an iris diaphragm, is demagnified by a 4-*f* telescopic array. A beam splitter (BS) directs a portion of this beam to a monitor camera, which is used to estimate the NA and the pupil intensity of the incident beam. A polarising beam splitter (PBS) and a quarter-wave plate (QWP) convert the polarisation of trapping light to circular polarisation, while also redirecting the backscattered light from a trapped nanoparticle to a split detector. A side-view image of the vacuum chamber, captured by an additional 4-*f* telescope, shows a silica nanoparticle trapped in the tweezers.

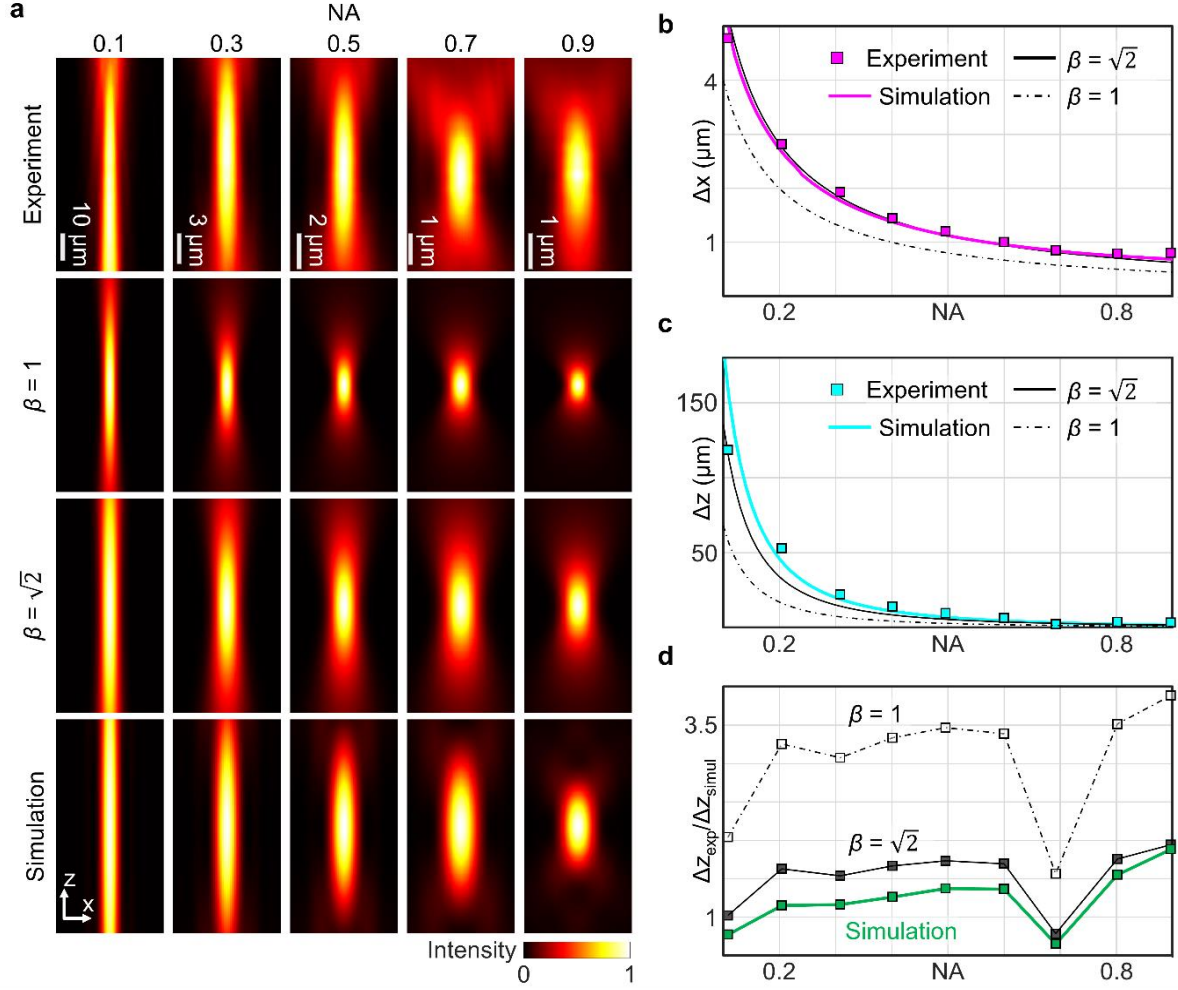

**Figure 2 | Analysis of point spread function (PSF)**

**a,** Comparison of the XZ cross-sections of the trapping light in the vicinity of focus. Shown are the experimentally determined PSF (first row), the predictions from the Gaussian beam model with $\beta = 1$ (second row), $\beta = \sqrt{2}$ (third row), and the full-field simulated PSF (fourth row). (**b, c**) Plots of (**b**) lateral ($\Delta x$) and (**c**) axial ($\Delta z$) full-width half-maxima (FWHMs) with respect to NA. **d,** The ratios of experimentally measured $\Delta z$ to the theoretically predicted values (black solid line: Gaussian beam with $\beta = \sqrt{2}$; black dashed line: Gaussian beam with $\beta = 1$; coloured solid line: simulation based on the angular spectrum decomposition method) as a function of NA.

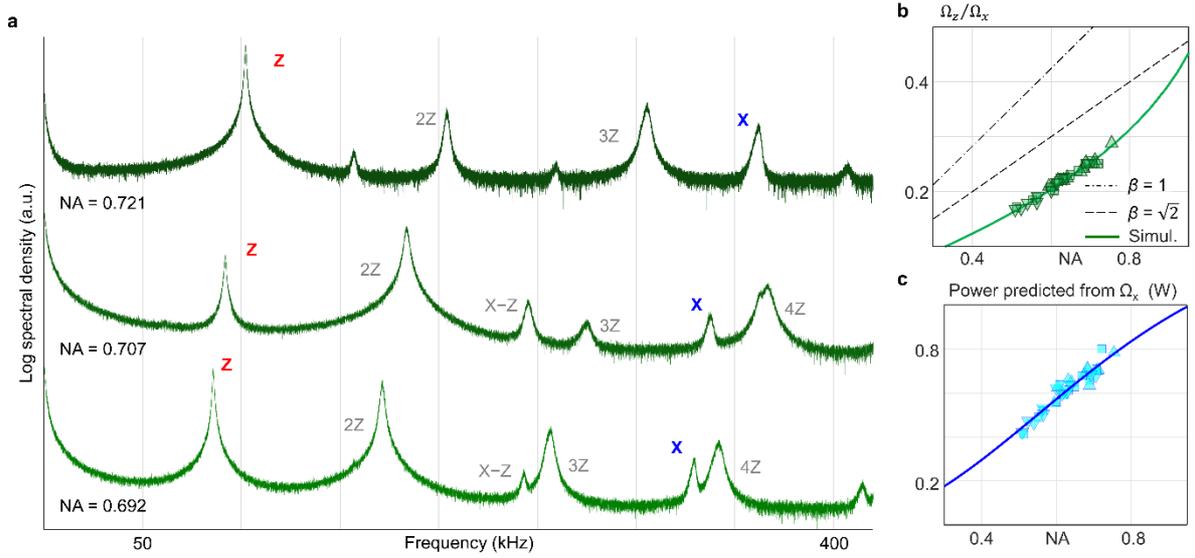

**Figure 3 | Analysis of trapping frequencies**

**a,** Representative data displaying the power spectral densities of particle's motion measured at different NAs. The data are median-filtered for improved visualization. **b,** The ratio between the axial and lateral trapping frequencies, $\Omega_z/\Omega_x$, plotted against NA. The solid coloured line represents the line predicted from the 3D simulation of the trapping light without any fitting parameters. The dashed black lines represent the predicted frequency ratios with the Gaussian beam model with different $\beta$. Three different particles were used in the experiments for reproducibility, each labelled as squares, upward triangles, and inverted triangles. **c,** The reconstructed laser focal powers derived from the full-field simulation with measured $\Omega_x$. The solid line represents the fitted power from Eq. 9, with $P(\text{NA}) = P_0 \cdot [1 - \exp(-\text{NA}^2/w_{NA}^2)]$, where $P_0$ is the only fitting parameter.

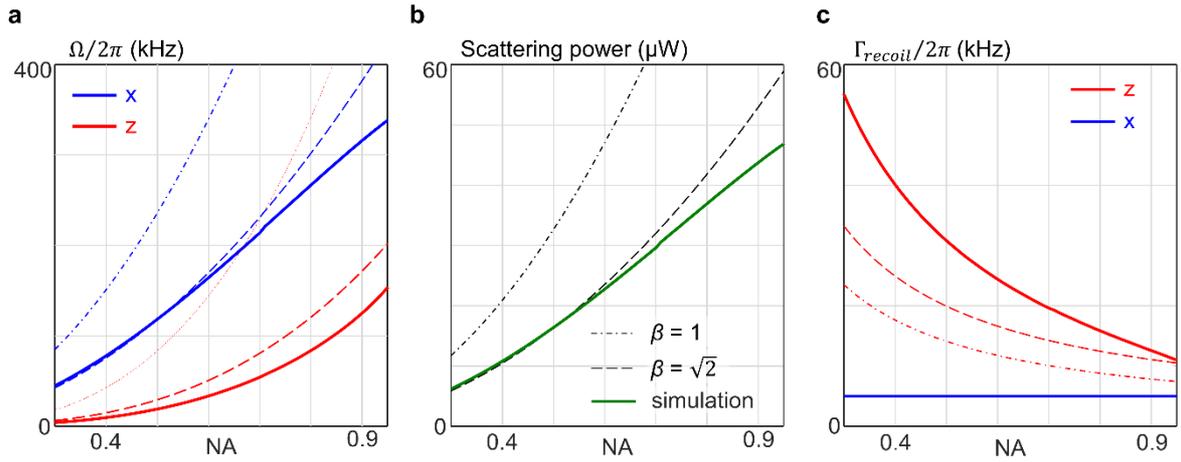

**Figure 4 | Predicted trap frequency, scattering power, and recoil heating when P = 250 mW.**

**a,** Predicted trapping frequencies in the lateral (*x*, blue lines) and axial (*z*, red lines) directions as functions of NA. **b,** Predicted scattering power versus NA. **c,** Predicted recoil heating rate along *x* (blue) and *z* (red) directions as functions of NA. The estimated recoil heating rate along the *x* direction is equivalent for both the Gaussian beam model and our simulation method. In all plots, solid lines represent the predictions from the full-field simulations, while dashed lines are obtained with the paraxial approximation with different $\beta$ values. All calculations assume $P$ = 250 mW.